\documentclass[sigconf,natbib=true]{acmart}
\AtBeginDocument{%
  }

\usepackage{algorithm}
\usepackage{algorithmic}
\usepackage{amsfonts}
\usepackage{amsmath}
\usepackage{booktabs}
\usepackage{array}
\newcolumntype{R}[1]{>{\raggedleft\arraybackslash}p{#1}}  
\usepackage{multirow}

\setcopyright{acmlicensed}
\copyrightyear{2018}
\acmYear{2018}
\acmDOI{XXXXXXX.XXXXXXX}
\acmConference[Conference acronym 'XX]{Make sure to enter the correct
  conference title from your rights confirmation email}{June 03--05,
  2018}{Woodstock, NY}
\acmISBN{978-1-4503-XXXX-X/2018/06}

\begin{document}

\title{Deliberate Latent Reasoning for Sequential Recommendation}

\author{Haobo Zhang}
\author{Yutao Zhu}
\affiliation{%
  \institution{Gaoling School of Artificial Intelligence}
  \institution{Renmin University of China}
  \city{Beijing}
  \country{China}
}
\email{{zhanghb,ytz}@ruc.edu.cn}

\author{Qiannan Zhu}
\affiliation{%
  \institution{School of Artificial Intelligence, Beijing Normal University;}
  \institution{Engineering Research Center of Intelligent Technology and Educational Application, MOE}
  \country{China}
}
\email{zhuqiannan@bnu.edu.cn}

\author{Zhicheng Dou}
\authornote{Corresponding author}
\affiliation{%
  \institution{Gaoling School of Artificial Intelligence}
  \institution{Renmin University of China}
  \city{Beijing}
  \country{China}}
\email{dou@ruc.edu.cn}

\renewcommand{\shortauthors}{Trovato et al.}

\begin{abstract}
Sequential recommenders are widely applied in real-world scenarios to model user preferences and behavior patterns based on user interactions. 
Recent advances in sequential recommendation enhance transformer-based models with latent reasoning to refine user representations by autoregressively generating intermediate reasoning vectors.
However, exploring the optimal reasoning depth for diverse user histories remains challenging: existing methods typically employ a fixed number of reasoning steps for all users. This ``one-size-fits-all'' strategy often leads to suboptimal representation learning due to over-thinking in simple histories or under-thinking in complex ones. 
To address this issue, we propose DESRec, a deliberate latent reasoning framework that performs personalized latent reasoning to learn more informative reasoning states and enhance user modeling for sequential recommendation.
Specifically, it designs a reasoning step selection module that identifies the optimal reasoning step for each user sequence during the training phase. Through a self-distillation mechanism, it compresses this most informative state into the final reasoning step, which serves as the final user representation for item ranking during inference.
Moreover, it introduces a progressive preference extraction mechanism for personalized reasoning, which dynamically segments a user’s history into a variable number of periods and summarizes each into a latent \textit{milestone} representing the user's preference for that period. It will be used as the input of the subsequent reasoning. The number of milestones varies with the users' history, enabling a personalized reasoning over users’ evolving interests.
Extensive experimental results on six public datasets and four backbone recommenders demonstrate the superiority of DESRec over eight baseline models.

\end{abstract}



\received{20 February 2007}
\received[revised]{12 March 2009}
\received[accepted]{5 June 2009}

\maketitle

\section{Introduction}
With the rapid growth of online information, recommender systems have become essential components of various platforms such as e-commerce, video, and music services~\cite{ecommerce2022, videorec2024, musicrec2024}.
As a key research direction, sequential recommendation mainly aims to capture users’ behavioral patterns by modeling their historical interactions.
Typical methods such as SASRec~\cite{SASRec2018} and BERT4Rec~\cite{BERT4Rec2019} mainly utilize transformer architectures to model user preferences.
These models represent user preferences using the last hidden state of the sequence and perform recommendations by matching them with candidate items. 
However, due to limitations in computational capacity and depth, such methods often cannot conduct fine-grained analysis over user sequences. 
As a result, they struggle to capture complex user interests, particularly for long-tail users and items.

\begin{figure}[t]
\centering
\includegraphics[width=0.92\linewidth]{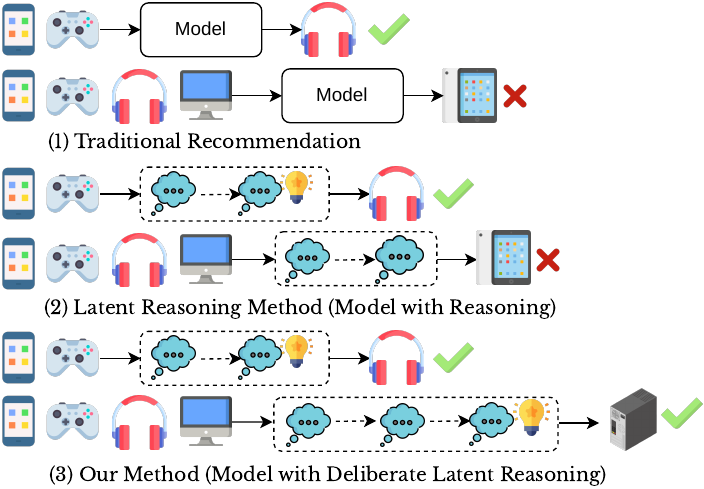}
\caption{Illustration of three recommendation paradigms.
(1) Traditional models directly perform recommendations without reasoning, failing to adequately analyze complex user histories.
(2) Latent reasoning models enhance recommendations by implicit reasoning, but apply a same reasoning strategy for all users, leading to inaccurate analysis under varying complexity.
(3) Our deliberate latent reasoning model learns the optimal reasoning state with different user histories, enabling more accurate reasoning and recommendations.}
\label{introduction_graph}
\end{figure}

Recently, researchers~\cite{Kimi2025, DeepSeekR1} have explored test-time computation to improve model performance. 
Chain-of-Thought~\cite{COT2022, COTtheory2023} first emerged as a prominent technique that enhances LLMs’ reasoning by decomposing complex tasks and thinking step-by-step.
Nevertheless, existing large reasoning models often adopt explicit reasoning, where reasoning steps are generated as explicit tokens~\cite{ReasonFlux2025, mathreason2024}. 
Unlike NLP tasks, where intermediate tokens are meaningful, recommenders focus on discrete IDs, which are difficult to supervise or explain in reasoning.
Moreover, mapping continuous embeddings to discrete IDs will introduce semantic loss~\cite{coconut2024}.
Therefore, recent work~\cite{TTCLatentreason2025, coconut2024} proposed latent reasoning in continuous embedding spaces without relying on explicit tokens.
This paradigm is more suitable for recommendation and enables better supervision of the reasoning process.
For example, ReaRec~\cite{ReaRec2025} autoregressively generates latent reasoning vectors as intermediate steps and performs recommendation by computing similarity between the final vector and candidates. It effectively improves the model's computation capacity and facilitates deeper analysis of user preferences.

While latent reasoning can effectively extend the computational capacity of models, there remain several limitations.
Current approaches typically \textbf{adopt the same reasoning strategy} (e.g., same number of reasoning steps) for all users.
However, users' preferences vary significantly across histories, and such a one-size-fits-all reasoning process fails to capture these differences. 
It may lead to over-reasoning for users with simple histories and introduce redundant signals; while for complex histories, it will result in under-reasoning and incomplete preference understanding.
Therefore, it is crucial to adjust the reasoning process to identify the optimal reasoning state and depth under different conditions.
As shown in Figure~\ref{introduction_graph}, the optimal reasoning steps vary considerably across users. 
For instance, in the Baby \& Products dataset, users exhibit highly diverse optimal reasoning depths.
A natural method is to dynamically select the reasoning depth at inference time. 
However, such test-time adaptation usually requires extra deployment-side judgments and computational cost, which complicates serving and may be unstable.
Therefore, the critical challenge lies in how to exploit the heterogeneity of optimal reasoning depths during training and learn more informative reasoning representations, while keeping inference simple with a fixed reasoning length.

\begin{figure}[t]
\centering
\includegraphics[width=0.95\linewidth]{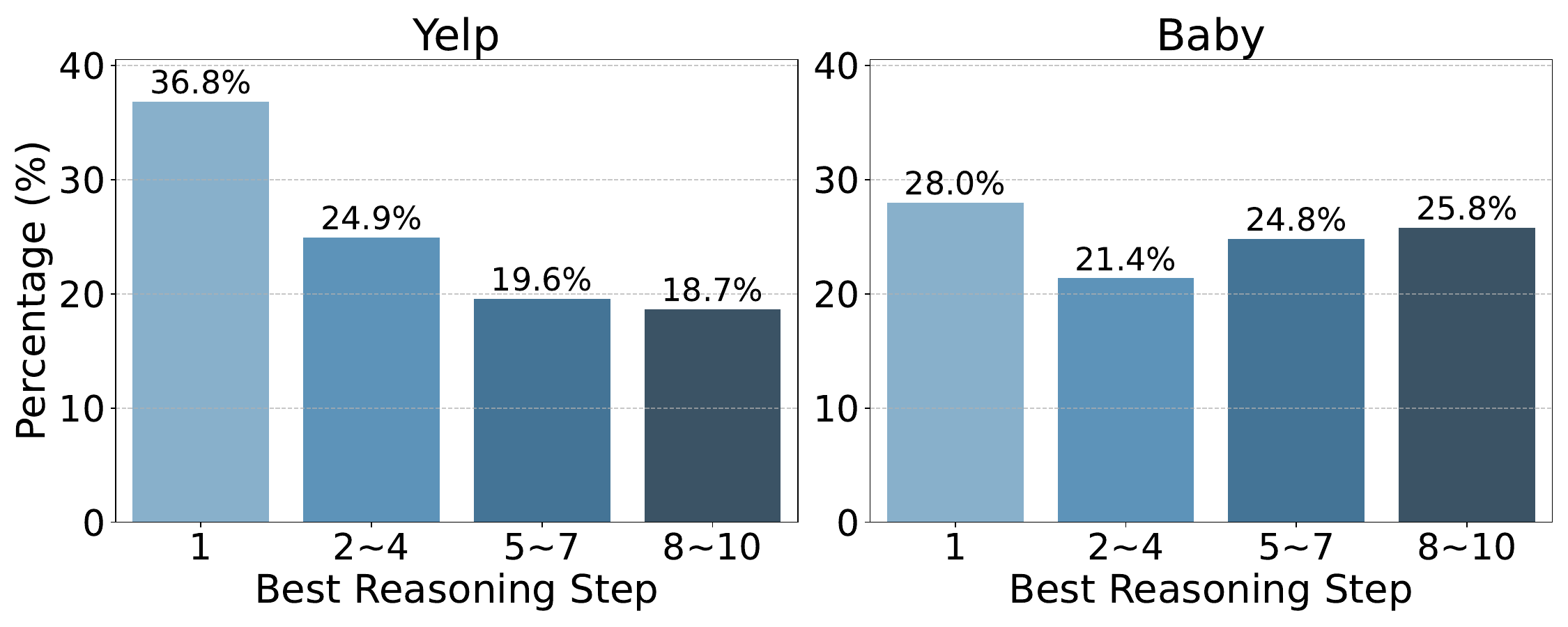}
\caption{Distribution of the optimal reasoning steps in two datasets. Each bar represents the percentage of samples whose optimal step is in the corresponding step range. A random subset from the 
test dataset is selected for this analysis. 
We train a basic latent reasoning model on SASRec with 1-10 reasoning steps, test each sample using them, and select the step with the highest NDCG score as the optimal step.
}
\label{introduction_graph}
\end{figure}

To address these challenges, we believe that models should carefully analyze the evolving user interests reflected in histories and learn to \textbf{identify and preserve the most informative reasoning state} under different conditions for better user preference modeling and reasoning.
Therefore, we propose DESRec, a deliberate latent reasoning framework for sequential recommendation. 
It analyzes the evolving user interests in histories and performs personalized latent reasoning that preserves the most informative reasoning state to enhance user preference modeling.
Specifically, we first design a reasoning step selection module (RESS) that identifies the optimal reasoning depth for each user during training by selecting the step with the highest recommendation score.
A self-distillation mechanism is then applied to preserve and compress critical information from the optimal reasoning step into the final-step representation, which serves as the final user representation for item ranking during inference.
Moreover, we introduce a progressive preference extraction module (PPEM) to model users' evolving interests and support more personalized reasoning. It dynamically segments each user's interaction history into different periods and summarizes each into a latent milestone.
The milestones are progressively integrated into the reasoning process, enabling the model to focus on distinct preference phases for personalized user modeling and better reasoning state learning.
By jointly modeling user preference evolution and exploiting optimal reasoning state and depth, these components construct a deliberate and personalized latent reasoning framework, leading to more accurate recommendations.

In summary, our main contributions are as follows:
\begin{itemize}
    \item we propose DESRec, a deliberate latent reasoning framework for sequential recommendation that enhances user preference modeling and learns more informative reasoning states by performing personalized latent reasoning.
    \item We design a reasoning step selection module that identifies each user's optimal reasoning step during training and compresses its information into the final-step representation via self-distillation, enabling effective recommendation with the final step at inference.
    \item We introduce a progressive preference extraction module that summarizes users' histories into a variable number of milestones and guides personalized reasoning to capture evolving preferences.
\end{itemize}

\section{Related Work}
\subsection{Sequential Recommendation}
Sequential recommendation mainly aims to model user preferences from interaction sequences for next-item prediction. 
Early methods~\cite{FPMC2010, Fossil2016} primarily rely on Markov chains to estimate item-item transitions using matrix factorization.
With the rise of deep learning, various architectures have been explored for sequence modeling, including CNN~\cite{CosRec2019, CGNIR2019}, RNN~\cite{GRU4Rec2016, HRNN2017}, and GNN~\cite{SRGNN2019, SURGE2021}.
For instance, Caser~\cite{Caser2018} employs CNNs to capture sequential patterns and model short- and long-term preferences.
Later, the advent of attention mechanisms and transformer architectures brought significant advances to sequential recommendation~\cite{SASRec2018, BERT4Rec2019, STAMP2018}. For example, 
S3-Rec~\cite{S3Rec2020} designs multiple self-supervised tasks to improve the representation learning of recommenders.
To alleviate cold-start issues, recent methods~\cite{MoRec2023, Recformer2023, KGIE2021} attempted to incorporate additional signals such as text content to enrich item representations. 
For example, UniSRec~\cite{UniSRec2022} leverages multi-domain data and item text embeddings to train the sequential model. 
However, most existing models cannot reason over user behavior sequences, limiting their ability to capture complex preferences. 
In this work, we introduce deliberate multi-step latent reasoning into sequential recommendation for better modeling of user preferences.

\subsection{Reasoning Models}
Reasoning is an essential capability to analyze complex problems and draw accurate conclusions. Early studies~\cite{COT2022, InstructRec2023} introduced Chain-of-Thought to encourage models to decompose problems into intermediate reasoning steps.
Recent work on test-time computation~\cite{BOT2024, TTCScale2024, GOT2024} has led to reasoning models such as OpenAI-o1~\cite{openaio12024} and DeepSeek-R1~\cite{DeepSeekR1}, which perform test-time reasoning by incorporating long COT prompting and RL strategies such as GRPO, achieving better performance on challenging problems.
Despite these strengths, these methods still suffer from the difficulties of supervising intermediate steps, inefficiency from long outputs, and semantic loss when mapping embeddings to discrete tokens.
To address these issues, researchers~\cite{coconut2024, VO-VAE2025, CCOT2024} have explored latent reasoning, where models perform multi-step reasoning in the latent space without generating explicit tokens. 
For instance, CODI~\cite{CODI2025} compresses natural language CoT into latent spaces via self-distillation for efficient reasoning.
In recommendation, ReaRec~\cite{ReaRec2025} applies latent reasoning to sequential recommenders by training transformer-based models to autoregressively generate latent reasoning vectors.
However, it does not explicitly exploit the varying reasoning requirements across users.

\begin{figure*}[t]
\centering
\includegraphics[width=0.98\linewidth]{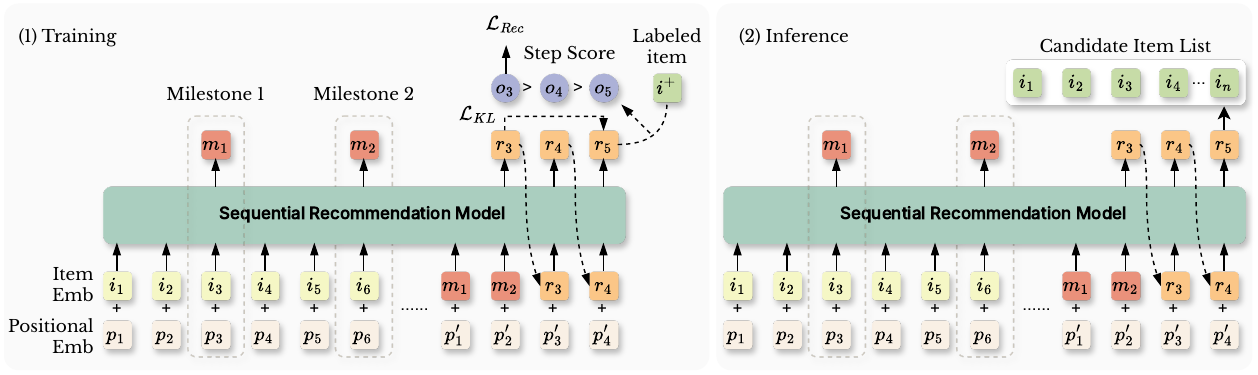}
\caption{The overall architecture of our DESRec model. }
\label{model_graph}
\end{figure*}

\section{Method}
In this section, we introduce DESRec, a deliberate latent reasoning framework designed to enhance sequential recommendation through personalized latent reasoning. The overall architecture is shown in Figure~\ref{model_graph}. 
We will first define the sequential recommendation task and formalize the latent reasoning process. 
Then, we introduce our optimal reasoning step selection and self-distillation mechanism to explore and preserve the most informative reasoning state for each user.
Last, we propose a progressive preference extraction method for capturing the evolving user interests and supporting personalized reasoning.

\subsection{Preliminaries}
\subsubsection{Problem definition}
Formally, we denote the set of users and items in the recommendation system as $U$ and $I$, with the size of the user and item sets as $N_U$ and $N_I$. 
The chronological user behavior sequence is denoted as $S_u = \{i_{1},...,i_{n}\}, \forall u \in U$. Each item is assigned a unique ID $i$ and has a set of textual attributes including titles, descriptions, and other attributes. The input texts for $i$ are constructed by concatenating its textual attributes into a sentence: $A_i = [a_1^i;...; a_m^i]$.  

For the sequential recommendation task, it takes the user $u$ and the user's behavior sequence $S_u$ as input, and predicts the next item $i_{n+1}$ the user is most likely to interact with. 
The sequential recommender will first transfer $S_u$ into a sequence of embeddings, and then encode them into a hidden state as the user representation $\mathbf{e}_u$. 
The likelihood of user $u$ interacting with the candidate item $i \in I$ is then computed based on the cosine similarity of their representations.
The training objective is to maximize the likelihood of the next item $i_{n+1}$, and formally it can be defined as:
\begin{equation}
    \max_{\Theta}\sum_{t=1}^{n - 1} P(i_{t+1} \mid S_{1:t}; \Theta),
\end{equation}
where $\Theta$ denotes the parameters of the recommender.

\subsection{Latent Reasoning Framework}
Traditional sequential recommendation models directly use the last hidden state of the user sequence for prediction, which limits the modeling of complex behaviors. 
To overcome this, researchers propose latent reasoning methods by performing multi-step inference in the latent space without generating explicit reasoning tokens.
At each step, the model autoregressively generates latent representations based on the previous reasoning state, enabling a more detailed understanding of user preferences.
In the following, we will take the transformer-based backbone SASRec as an example and use it to encode the input interaction sequence. 
Formally, given a user's interaction sequence $S_u$, each item $i_t$ in the sequence is first mapped to the item embedding $\mathbf{e}_t$ through the embedding matrix $\mathbf{E}$, and its position is encoded as $\mathbf{p}_t$. 
The input representation sequence can be defined as:
\begin{equation}
\mathbf{X}^{(0)} = [\mathbf{e}_{1} + \mathbf{p}_1, \mathbf{e}_{2} + \mathbf{p}_2, \dots, \mathbf{e}_{n} + \mathbf{p}_{n}],
\end{equation}
where $\mathbf{p}_t$ derives from a learnable positional embedding matrix $\mathbf{P}$. This sequence is passed into a transformer encoder $f$ to produce contextualized hidden states:
\begin{equation}
\mathbf{H}^{(0)} = f(\mathbf{X}^{(0)}),
\end{equation}
where $\mathbf{H}^{(0)} \in \mathbb{R}^{n\times d}$ denotes the output hidden state without reasoning, $f$ consists of some stacked multi-head attention and FFN layers. The hidden state $\mathbf{h}_{n}^{(0)}$ at the last position is used as the initial reasoning representation $\mathbf{r}_1$.
To perform multi-step latent reasoning, the model autoregressively generates the next latent representation from $\mathbf{h}_{n+1}^{(1)}$.
To distinguish the item sequence from the reasoning steps, we define another positional embedding matrix $\mathbf{P}'$ for the reasoning steps. 
At step $s$, the previous reasoning output $\mathbf{h}_{n+s-1}^{(s-1)}$ is appended to the input sequence with its corresponding positional embedding:
\begin{equation}
\mathbf{X}^{(s)} = [\mathbf{X}^{(s-1)}; \mathbf{h}_{n+s-1}^{(s-1)} + \mathbf{p}_{s}'].
\end{equation}
We then use the transformer to encode this sequence and take the last hidden state $\mathbf{h}_{n+s}^{(s)}$ as the current reasoning state $\mathbf{r}_{s+1}$:
\begin{equation}
\mathbf{H}^{(s)} = f(\mathbf{X}^{(s)}), \: \mathbf{r}_{s+1} = \mathbf{h}_{n+s}^{(s)} = \mathbf{H}^{(s)}[-1].  
\end{equation}
This process will continue for a fixed number of steps $T$ and get the final reasoning state $\mathbf{h}_{{n}+T}^{(T)}$. It will be used to predict the next item via cosine similarity with item embeddings. The training objective is to maximize the log-likelihood of the ground-truth next item $i^+$:
\begin{equation}
\mathcal{L} = -\log \frac{\exp(\mathbf{h}_{n+T}^{(T)} \cdot \mathbf{e}_{i^+})}{\sum_{i \in \mathcal{I}} \exp(\mathbf{h}_{n+T}^{(T)} \cdot \mathbf{e}_i)}.
\end{equation}

\subsection{Personalized Training Strategy via Optimal Step Selection and Distillation}~\label{sec:RESS}

While latent reasoning enhances the user preference modeling, a fixed number of reasoning steps does not benefit all users equally.
A static reasoning strategy often causes over-reasoning or under-reasoning for users with simple or complex histories, leading to suboptimal reasoning.
Therefore, we propose a reasoning step selection mechanism (RESS) that dynamically identifies the optimal reasoning step for each user during training and distills the information into the last step. It enables personalized reasoning during the training phase, while retaining efficient fixed-step inference at test time.

\subsubsection{Optimal Reasoning Step Selection}~\label{sec:step_select}
Let $\{\mathbf{r}_1, ..., \mathbf{r}_T\}$ denote the latent reasoning vectors at each step for user $u$. To determine the most informative reasoning step, we compute the similarity score for each step $s$ using the reasoning embedding $\mathbf{r}_s$ and the embedding of the ground-truth item $i^+$.:
\begin{equation}
o_s =  \text{sim}(\mathbf{r}_s, \mathbf{e}_{i^+}),
\end{equation}
where $\text{sim}(\cdot)$ denotes the cosine similarity, and $\mathbf{e}_{i^+}$ is the embedding of the target item. The optimal step $s^*$ is then defined as the step with the highest similarity score:
\begin{equation}
s^* = \arg\max_{1 \leq s \leq T} o_s,
\end{equation}

\subsubsection{Self-Distillation}
After identifying the optimal reasoning step, we adopt a self-distillation mechanism to compress its information into the last step $\mathbf{r}_T$. 
Specifically, the most optimal step serves as a teacher to supervise the representation learning of the last step, allowing the model to preserve critical signals from all reasoning steps without losing early but informative cues.
It enables highly efficient and effective at test time by learning to capture the most informative reasoning states and implicitly retaining the benefits of multi-step reasoning.

Formally, we denote $\mathbf{r}_{s*}$ as the reasoning embedding at the optimal step $s^*$ for user $u$, and $\mathbf{r}_T$ as the final-step reasoning embedding. 
We first compute the scores between the reasoning embeddings and all candidate items. For the optimal step $s^*$, we calculate its score with each item $i$ as:
\begin{equation}
P_{s^*}(i|u) = \frac{\exp(\text{sim}(\mathbf{r}_{s^*}, \mathbf{e}_{i})/\tau)}{\sum_{j \in I} \exp(\text{sim}(\mathbf{r}_{s^*}, \mathbf{e}_{j})/\tau))},
\end{equation}
where $\mathbf{e}_{i}$ is the embedding of item $i$, and $\tau$ is a temperature hyperparameter. The resulting probability distribution over all items is then given by:
\begin{equation}
P_{s^*}(I|u) = [P_{s^*}(i_1|u), P_{s^*}(i_2|u),..., P_{s^*}(i_{N_I}|u)].
\end{equation}

Similarly, we compute the scores between the final reasoning step $T$ and all candidate items:
\begin{equation}
P_{T}(i|u) = \frac{\exp(\text{sim}(\mathbf{r}_{T}, \mathbf{e}_{i})/\tau)}{\sum_{j \in I} \exp(\text{sim}(\mathbf{r}_{T}, \mathbf{e}_{j})/\tau))}.
\end{equation}

The probability distribution over all items is given by:
\begin{equation}
P_{T}(I|u) = [P_{T}(i_1|u), P_{T}(i_2|u),..., P_{T}(i_{N_I}|u)].
\end{equation}

Finally, we use the distribution $P_{s^*}(I|u)$ as the teacher to guide the learning of $P_{T}(I|u)$ by minimizing the Kullback-Leibler (KL) divergence:
\begin{equation}\label{KL_loss}
\mathcal{L}_{\text{KL}} = \text{KL}\left(P_{s^*}(I|u) \, \| \, P_{T}(I|u)\right).
\end{equation}

\subsection{Progressive Preference Extraction for Personalized Reasoning}~\label{sec:PPEM}
While multi-step latent reasoning refines user representations through iterative reasoning, it may still struggle to capture evolving preferences in long or complex interaction sequences.
Therefore, we propose a progressive preference extraction mechanism (PPEM) for more personalized latent reasoning.
It partitions a user's interaction history into different periods and summarizes each period into a latent milestone, where the number of milestones naturally scales with the user's history length.
The milestones capture the evolving user interests, and are progressively integrated into the reasoning process.
By reasoning over the milestones, the model can better focus on distinct preference phases, achieving more personalized user preference modeling.

\subsubsection{Structured Preference Extraction}
This component is designed to capture the nuanced user preferences in different periods. It extracts preference representations from different periods into some milestones.
We apply a sliding window of fixed length over the historical sequence to capture the preference representations, which reflect the user's evolving interests in different temporal phases.

Formally, given a user’s encoded sequence output from the transformer $\mathbf{H}=[\mathbf{h}_1, \mathbf{h}_2, ..., \mathbf{h}_{n}]$, where $\mathbf{h}_t$ is the hidden state at position $t$, we apply a sliding window of fixed size $w$ and stride $D$ to partition the sequence into segments:
\begin{equation}
\mathcal{W}_j = \{\mathbf{h}_{(j-1)D+1}, ..., \mathbf{h}_{(j-1)D+w}\}, \quad j = 1, 2, ..., M
\end{equation}
where $M = \left\lfloor \frac{n - w}{D} \right\rfloor + 1$ is the number of windows.

For each window $\mathcal{W}_j$, we select the hidden state at the last position as the \textit{milestone} for that period:
\begin{equation}~\label{last_pool}
\mathbf{m}_j = \mathbf{h}_{(j-1)D + w}.
\end{equation}

The resulting set of the milestones is denoted as $\mathbf{M} = [\mathbf{m}_1, \mathbf{m}_2, ..., \\ \mathbf{m}_M] \in \mathbb{R}^{M \times d}$. These milestones provide a structured view of the user's preferences at different periods, which will be used in subsequent latent reasoning.

\subsubsection{Progressive Reasoning}
To better capture the evolving nature of user preferences over time, we introduce a progressive reasoning strategy that incorporates the milestones into the latent reasoning process. This component guides the model to reason in a temporally structured manner based on structured preference representations.

Formally, let $[\mathbf{e}_1, ..., \mathbf{e}_{n}]$ denote the original item sequence embeddings, and $\{\mathbf{m}_1, ..., \mathbf{m}_M\}$ represent the milestones extracted from the sliding windows over the interaction history. The initial input for the first reasoning step is:
\begin{equation}
\mathbf{X}_r^{(1)} = [\mathbf{e}_1 + \mathbf{p}_1, ..., \mathbf{e}_{n} + \mathbf{p}_{n}, \mathbf{m}_1 + \mathbf{p}_1', ..., \mathbf{m}_M + \mathbf{p}_M'],
\end{equation}
where $\mathbf{p}_j$ and $\mathbf{p}_j'$ are the positional embeddings for items and reasoning steps to distinguish the two types of inputs.

Following this initialization, the model autoregressively performs $T$ steps of latent reasoning. At each step $s$, the model takes the concatenation of the previous sequence $\mathbf{X}_r^{(s-1)}$ and the last reason state $\mathbf{h}^{(s-1)}_{n+M+s-1} + \mathbf{p}'_{M+s}$ as input and produces a latent reasoning representation $\mathbf{r}_{s+M+1} = \mathbf{h}^{(s)}_{n+M+s}$:
\begin{equation}
\begin{aligned}
\mathbf{X}_r^{(s)} &= [\mathbf{X}_r^{(s-1)}, \mathbf{h}^{(s-1)}_{n+M+s-1} + \mathbf{p}'_{M+s}], \\
\mathbf{H}^{(s)} &= f(\mathbf{X}_r^{(s)}),\: \mathbf{h}^{(s)}_{n+M+s} = \mathbf{H}^{(s)}[-1].
\end{aligned}
\end{equation}

The representation at the final step $\mathbf{h}^{(T)}_{n+M+T}$ is used to compute the prediction score for the next item.

\subsubsection{Training Objective}
In addition to the distillation loss, we also apply a standard recommendation loss (e.g., cross-entropy) between the optimal step and the ground-truth item:
\begin{equation}
\mathcal{L}_{\text{rec}} = -\log P_{s^*}(i^+ | u).
\end{equation}

We train the model by minimizing the following loss:
\begin{equation}
\mathcal{L} = \mathcal{L}_{\text{rec}} + \lambda \mathcal{L}_{\text{KL}},
\end{equation}
where $\lambda$ is a hyperparameter balancing the two objectives.

After training, the model learns to retain the most critical reasoning knowledge in the last step.  
For inference, we directly use the last reasoning embedding $\mathbf{r}_T$ to compute the ranking scores over all items for recommendation:
\begin{equation}
\hat{y}_{u,i} = \text{sim}(\mathbf{r}_T, \mathbf{e}_i), \quad \forall i \in I.
\end{equation}

\subsection{Time Complexity Analysis}

We now analyze the time complexity of DESRec framework, taking into account both the base encoder and the reasoning module.

\textbf{Base Encoding.} Let $C$ denote the length of the input sequence and $d$ the hidden dimension. For a transformer-based encoder with $L$ layers, each layer involves a multi-head self-attention (MHSA) layer and a feed-forward network (FFN). The MHSA module has a complexity of $\mathcal{O}(C^2d + Cd^2)$, and the FFN's complexity is $\mathcal{O}(Cd^2)$. Thus, the overall cost of the base encoder is $\mathcal{O}(L(C^2d + Cd^2))$.

\textbf{Reasoning Extension.} In the reasoning process, we first generate $M$ milestones and then sequentially generate $T$ reasoning steps and append them to the sequence. To avoid recomputing attention for the full sequence at each step, we cache previous key-value pairs. 
At the first step, we need to compute the attention between the milestones and the previous sequences, and the cost is $\mathcal{O}(L(CMd+Md^2))$.
At step $s$, we only compute attention between the new step and the previous $C+M+s-1$ step, and the cost is $\mathcal{O}(L((C+M+s-1)d + d^2))$. The total complexity of $T$ reasoning steps are $\mathcal{O}(L(T(C+M)d + T^2 d + T d^2))$.

In most practical settings, $T$ and $M$ is much smaller than $C$ (e.g., $T \ll C$, $M = C/n, n\approx10$), allowing the overhead to be approximated as $\mathcal{O}(L(C^2d + Cd^2)/n)$ and $\mathcal{O}(L(TCd + Td^2))$. Therefore, the reasoning mechanism introduces a moderate and manageable computational cost and maintains the scalability of the model.

\section{Experimental Setup}
\subsection{Dataset}

We select six real-world public datasets from Yelp, Amazon, and MovieLens-1M. For all datasets, ratings above 3 are treated as positive feedback, followed by k-core filtering and chronological train/valid/test splitting. Detailed statistics are shown in Table~\ref{dataset}.

(1) \textbf{Yelp}\footnote{\url{https://business.yelp.com/data/resources/open-dataset/}} is a business recommendation dataset. Following~\cite{ReaRec2025}, we apply 20-core filtering and use the item name, location (city and state), and categories for textual encoding. Two cutoff dates for splitting the datasets are September 4, 2018, and May 13, 2020.

(2) \textbf{Amazon}\footnote{\url{https://amazon-reviews-2023.github.io/}} is a product recommendation dataset that contains user-product interactions and metadata. We utilized four subsets: Video \& Games, Software, CDs \& Vinyl, and Baby \& Products. Following~\cite{ReaRec2025}, we apply 5-core filtering for users and use the product attributes of title, description, and price for textual encoding. The datasets are split based on the official absolute timestamps\footnote{\url{https://amazon-reviews-2023.github.io/data_processing/5core.html}}.

(3) \textbf{MovieLens-1M}\footnote{\url{https://grouplens.org/datasets/movielens/1m/}} is a movie recommendation dataset. We apply 20-core filtering to the dataset and use the movie's title and genres for textual encoding. Two cutoff timestamp for splitting the datasets are 1028504458 and 1037479524, corresponding to 80\% and 90\% of the timeline, respectively.

\begin{table}[t]
\small
\centering
\caption{Statistics of the datasets.}
\begin{tabular}{p{.75cm} R{.8cm} R{.92cm} R{.87cm} R{.92cm} R{.87cm}R{.95cm}}
\toprule 
Dataset & Yelp & Video \& \newline Games & Software & CDs \& \newline Vinyl & Baby \& \newline Products & MovieLens-1M \\
\midrule
\#Users & 13,083 & 54,001 & 89,280 & 89,590 & 140,292 & 6,040 \\
\#Items & 10,697 & 22,735 & 17,233 & 88,462 & 30,689 & 3,883 \\
\#Inter & 509,212 & 519,505 & 831,136 & 1,240,407 & 818,207  & 1,000,209 \\
\#Inter/u & 38.92 & 9.62 & 9.31 & 13.85 & 5.83 & 165.60  \\
\#Inter/i & 47.60 & 22.85 & 48.23 & 14.02 & 26.66 & 257.59  \\
Sparisty & 99.64\% & 99.96\% & 99.95\% & 99.98\% & 99.98\% & 95.75\%\\
\bottomrule
\end{tabular}
\label{dataset}
\end{table}

\subsection{Backbone Models and Baselines}

To comprehensively evaluate our framework, we consider both ID-based and text-based sequential recommendation models as backbones. Following the baseline settings in~\cite{ReaRec2025}, we adopt SASRec and BERT4Rec as ID-based models, and UniSRec and MoRec as text-based models.

$\bullet$ \textbf{SASRec}~\cite{SASRec2018}: It is a Transformer-based sequential recommender that models user behavior patterns from item ID sequences using self-attention.

$\bullet$ \textbf{BERT4Rec}~\cite{BERT4Rec2019}: It adopts a bidirectional Transformer with masked item prediction to learn context-aware sequential user preferences.

$\bullet$ \textbf{UniSRec}~\cite{UniSRec2022}: It learns universal item and sequence representations from textual item embeddings for cross-domain sequential recommendation.

$\bullet$ \textbf{MoRec}~\cite{MoRec2023}: It leverages pretrained textual and visual encoders to obtain multi-modal item representations and models user behavior sequences with a Transformer.

For the baseline models, we select a latent reasoning method:

$\bullet$ \textbf{ReaRec}~\cite{ReaRec2025}: It leverages the latent reasoning method to enhance the sequential recommendation. It proposes two variants: ReaRec-ERL and ReaRec-PRL. ERL employs ensemble learning to aggregate multi-step reasoning states and introduces a regularizer to mitigate homogenization. PRL adopts contrastive learning to improve robustness and incorporates temperature annealing to enhance sequential reasoning. We use both ERL and PRL as baselines.

To further verify the effectiveness of DESRec, we include two conventional sequential recommendation models, \textbf{Recformer} and \textbf{LRURec}, as well as an LLM-based reasoning method \textbf{COT-Rec}.

$\bullet$ \textbf{Recformer}~\cite{Recformer2023}: It represents item sequences with key-value attribute pairs and employs a Transformer to encode them.

$\bullet$ \textbf{LRURec}~\cite{LRURec2024}: It adopts a linear recurrent unit-based architecture to efficiently model long-range dependencies in user behavior sequences.

$\bullet$ \textbf{COT-Rec}~\cite{COTRec2025}: It leverages an LLM for user preference analysis and item perception analysis, and then conducts explicit COT reasoning to assess whether candidates satisfy user preferences for recommendation.

\subsection{Evaluation Metrics}
We use Recall@k and NDCG@k as our metrics to evaluate the effectiveness and performance of our model, which are widely used in the recommendation field~\cite{ReaRec2025, UniSRec2022}. We set k as 10 and 20 in the experiments. For a comprehensive evaluation, we adopt a full-ranking evaluation strategy, where each user’s candidate list includes all items for ranking. 
Following common practice, all evaluation metrics in the tables are reported after being multiplied by 100 for better readability.

\subsection{Implementation Details}
In our experiments, we set the embedding size as 256, the batch size as 512, and the learning rate as 0.0001. We use the Adam optimizer to optimize all the models. Following previous works~\cite{ReaRec2025}, we set the maximum of the history length for each sample as 50. The weight of self-distillation loss $\lambda$ is set to 1.0. The temperature $\tau$ is 1.0 and 0.5 for ID-based and text-based backbones, respectively. The window size $w$ and stride $D$ in progressive preference extraction are both set to 10. For the BERT4Rec backbone model, following previous works~\cite{ReaRec2025}, we adopt the bidirectional attention in the item sequence part and unidirectional attention in the reasoning steps. For the text-based backbone model, we leverage Llama-3.1-8B~\cite{llama3} as the text encoder to encode the item text contents. The average of the last hidden states is used as the text representation, and it is further reduced to 768 dimensions via Principal Component Analysis (PCA) to retain core semantic features.

\begin{table*}[t]
\small
\centering
\caption{The recommendation results of Recall and NDCG@10, 20 on four backbones and six datasets. The best and second-best results are shown in bold and underline, respectively. All values are multiplied by 100 when reporting. * indicates the model outperforms all baselines significantly with paired t-test for p\textless 0.05.}
\setlength{\tabcolsep}{2pt}{
\begin{tabular}{clcccccccccccccccc}
\toprule
\multirow{2}{*}{Dataset} & \multicolumn{1}{c}{\multirow{2}{*}{Method}} & \multicolumn{4}{c}{SASRec} & \multicolumn{4}{c}{BERT4Rec} & \multicolumn{4}{c}{UniSRec} & \multicolumn{4}{c}{MoRec} \\
\cmidrule(lr){3-6}\cmidrule(lr){7-10}\cmidrule(lr){11-14}\cmidrule(lr){15-18}    
  & & N@10 & N@20 & R@10 & R@20 & N@10 & N@20 & R@10 & R@20 & N@10 & N@20 & R@10 & R@20 & N@10 & N@20 & R@10 & R@20 \\
 \midrule
\multirow{5}{*}{Yelp} & Base & 3.54 & 4.47 & 6.32 & 10.15& 3.64 & 4.52 & 6.52 & 10.32 & 3.86 & 4.93 & 7.43 & 12.39 & 3.71 & 4.79 & 7.37 & 12.28 \\
 & ERL & 3.65 & 4.63 & 6.61 & 10.55 & 3.67 & \underline{4.72} & 6.79 & 10.73 & 4.08 & 5.28 & \underline{8.01} & 13.10 & 3.82 & 5.06 & \underline{7.78} & \underline{12.74} \\
 & PRL & \underline{3.68} & \underline{4.69} & \underline{6.79} & \underline{10.86} & \underline{3.70} & 4.71 & \underline{7.06} & \underline{10.89} & \underline{4.13} & \underline{5.36} & 0.0784 & \underline{13.12} & \underline{3.86} & \underline{5.10} & 7.66 & 12.63 \\
 & DESRec & \textbf{3.93}* & \textbf{4.99}* & \textbf{7.47}* & \textbf{11.65}* & \textbf{3.96}* & \textbf{4.93}* & \textbf{7.47}* & \textbf{11.35} & \textbf{4.49}* & \textbf{5.78}* & \textbf{8.55}* & \textbf{13.67} & \textbf{4.12}* & \textbf{5.42}* & \textbf{8.28}* & \textbf{13.47}* \\
 & Impr. & 6.79\% & 6.40\% & 10.01\% & 7.27\% &  7.03\% & 4.45\% & 5.81\% & 4.22\% & 8.72\% & 7.84\% & 6.74\% & 4.19\%  & 6.74\% & 6.27\% & 6.43\% & 5.73\% \\
 \midrule
\multirow{5}{*}{Video \& Games} & Base & 1.97 & 2.64 & 4.37 & 6.66 & 2.02 & 2.65 & 4.46 & 6.72 & 3.27 & 4.21 & 6.54 & 9.65 & 3.41 & 4.22 & 6.63 & 9.70 \\
 & ERL & \underline{2.20} & \underline{2.72} & \underline{4.63} & \underline{6.81} & \underline{2.22} & 2.73 & \underline{4.84} & \underline{6.99} & \underline{3.68} & \underline{4.54} & \underline{6.96} & \underline{10.28} & \underline{3.72} & \underline{4.57} & \underline{7.01} & \underline{10.19} \\
 & PRL & 2.13 & 2.69 & 4.49 & 6.72 & 2.20 & \underline{2.77} & 4.61 & 6.91 & 3.56 & 4.43 & 6.72 & 10.17 & 3.61 & 4.54 & 6.87 & 10.07 \\
 & DESRec & \textbf{2.33}* & \textbf{2.91}* & \textbf{4.70} & \textbf{7.02} & \textbf{2.44}* & \textbf{3.05}* & \textbf{5.08}* & \textbf{7.32}* & \textbf{3.90}* & \textbf{4.76} & \textbf{7.41}* & \textbf{10.83}* & \textbf{3.99}* & \textbf{4.81}* & \textbf{7.38} & \textbf{10.62} \\
 & Impr. & 5.91\% & 6.99\% & 1.51\% & 3.08\% & 9.91\% & 10.11\% & 4.96\% & 4.72\% & 5.98\% & 4.85\% & 6.47\% & 5.35\% & 7.26\% & 5.25\% & 5.28\% & 4.22\%\\
 \midrule
\multirow{5}{*}{Software} & Base & 7.96 & 10.14 & 15.65 & 23.98 & 8.14 & 10.26 & 16.03 & 24.06 & 8.62 & 10.52 & 16.27 & 24.36 & 8.72 & 10.64 & 16.31 & 24.64 \\
 & ERL & 8.32 & 10.28 & \underline{16.16} & \underline{24.02} & 8.36 & 10.34 & \underline{16.31} & 24.08 & 9.02 & 11.09 & \underline{16.94} & \underline{25.48} & 9.15 & \underline{11.38} & \underline{17.04} & \underline{25.45} \\
 & PRL & \underline{8.64} & \underline{10.54} & 15.97 & 23.93 & \underline{8.66} & \underline{10.61} & 15.99 & \underline{24.37} & \underline{9.11} & \underline{11.12} & 16.87 & 25.22 & \underline{9.25} & 11.26 & 16.94 & 24.94 \\
 & DESRec & \textbf{9.14}* & \textbf{11.24}* & \textbf{16.78} & \textbf{25.07}* & \textbf{9.12}* & \textbf{11.09}* & \textbf{17.13}* & \textbf{25.09} & \textbf{9.41}* & \textbf{11.64}* & \textbf{17.51}* & \textbf{26.42} & \textbf{9.63}* & \textbf{11.83}* & \textbf{17.71}* & \textbf{26.46}* \\
 & Impr. & 5.79\% & 6.64\% & 3.84\% & 4.37\% & 5.31\% & 4.52\% & 5.03\% & 2.95\% & 3.29\% & 4.68\% & 3.36\% & 3.69\% &  4.11\% & 3.95\% & 3.93\% & 3.97\% \\
 \midrule
\multirow{5}{*}{CDs \& Vinyl} & Base & 1.81 & 2.18 & 3.96 & 5.41 & 1.89 & 2.39 & 4.08 & 6.04 & 2.20 & 2.81 & 4.79 & 7.01 & 2.13 & 2.73 & 4.22 & 6.40 \\
 & ERL & \underline{2.08} & \underline{2.44} & \underline{4.50} & \underline{5.91} & \underline{2.12} & 2.52 & \underline{4.59} & 6.17 & \underline{2.36 }& \underline{3.06} & \underline{4.99} & \underline{7.34} & 2.21 & 2.85 & 4.46 & 6.61 \\
 & PRL & 2.01 & 2.41 & 4.31 & 5.89 & 2.11 & \underline{2.58} & 4.53 & \underline{6.39} & 2.31 & 2.96 & 4.89 & 7.17 & \underline{2.28} & \underline{2.91} & \underline{4.63} & \underline{6.75} \\
 & DESRec & \textbf{2.37}* & \textbf{2.76}* & \textbf{5.15}* & \textbf{6.70}* & \textbf{2.47}* & \textbf{2.89}* & \textbf{5.31}* & \textbf{6.95}* & \textbf{2.59}* & \textbf{3.26}* & \textbf{5.22}* & \textbf{7.89} & \textbf{2.56}* & \textbf{3.21}* & \textbf{5.01}* & \textbf{7.53}* \\
 & Impr. & 13.94\% & 13.11\% & 14.44\% & 13.37\% &  16.51\% & 12.02\% & 15.69\% & 8.76\% & 9.75\% & 6.54\% & 4.61\% & 7.49\% & 12.28\% & 10.31\% & 8.21\% & 11.56\%  \\
 \midrule
\multirow{5}{*}{Baby \& Products} & Base & 1.06 & 1.46 & 2.21 & 3.91 & 1.08 & 1.49 & 2.41 & 4.02 & 1.62 & 2.17 & 3.25 & 5.42 & 1.52 & 2.06 & 3.09 & 5.13 \\
 & ERL & 1.16 & 1.53 & 2.34 & 4.02 & 1.17 & 1.61 & 2.43 & \underline{4.27} & 1.73 & 2.24 & 3.38 & 5.69 & 1.61 & 2.18 & 3.18 & 5.31 \\
 & PRL & \underline{1.20} & \underline{1.62} & \underline{2.37} & \underline{4.03} & \underline{1.22} & \underline{1.65} & \underline{2.49} & 4.22 & \underline{1.83} & \underline{2.40} & \underline{3.63} & \underline{5.80} & \underline{1.67} & \underline{2.22} & \underline{3.25} & \underline{5.46} \\
 & DESRec & \textbf{1.36}* & \textbf{1.78}* & \textbf{2.61}* & \textbf{4.26} & \textbf{1.39}* & \textbf{1.82}* & \textbf{2.70}* & \textbf{4.39} & \textbf{2.00}* & \textbf{2.56}* & \textbf{3.89}* & \textbf{6.09}* & \textbf{1.84}* & \textbf{2.40}* & \textbf{3.63}* & \textbf{5.86}* \\
 & Impr. & 13.33\% & 9.88\% & 10.13\% & 5.71\% &  13.93\% & 10.30\% & 8.43\% & 2.81\% & 9.29\% & 6.67\% & 7.16\% & 5.00\% & 10.18\% & 8.11\% & 11.69\% & 7.33\% \\
 \midrule
\multirow{5}{*}{MovieLens-1M} & Base & 6.07 & 7.52 & 11.84 & 17.86 & 5.98 & 7.30 & 11.68 & 17.46 & 6.44 & 8.06 & 12.67 & 19.04 & 6.30 & 7.82 & 12.33 & 18.39\\
 & ERL & 6.54 & \underline{8.12} & 12.09 & 18.62 & 6.41 & 8.02 & \underline{11.98} & 18.25 & 7.01 & 8.65 & \underline{13.57} & \underline{20.05} & \underline{6.72} & 8.33 & \underline{12.86} & 19.63 \\
 & PRL & \underline{6.48} & 8.09 & \underline{12.13} & \underline{18.79} & \underline{6.43} & \underline{8.06} & 11.94 & \underline{18.31} & \underline{7.04} & \underline{8.68} & 13.40 & 19.96 & 6.66 & \underline{8.50} & 12.84 & \underline{19.78}\\
 & DESRec & \textbf{6.89}* & \textbf{8.66}* & \textbf{12.81}* & \textbf{19.83}* & \textbf{6.85}* & \textbf{8.47}* & \textbf{12.57}* & \textbf{19.31}* & \textbf{7.35}* & \textbf{9.13}* & \textbf{14.36}* & \textbf{20.94} & \textbf{7.18}* & \textbf{8.95}* & \textbf{13.71}* & \textbf{20.63}\\
 & Impr. & 5.35\% & 6.65\% & 5.61\% & 5.53\% & 6.53\% & 5.09\% & 4.92\% & 5.46\% & 4.40\% & 5.18\% & 5.82\% & 4.44\% & 6.85\% & 5.29\% & 6.61\% & 4.30\% \\
 \bottomrule
\end{tabular}
}
\label{IDmodel_result}
\end{table*}

\begin{table}[t]
\small
\centering
\caption{The results of Recall and NDCG@20 on three datasets for conventional recommenders and LLM-based reasoning methods. The best results are shown in bold. All values are multiplied by 100 when reporting.}
\begin{tabular}{lcccccc}
\toprule
 \multicolumn{1}{c}{\multirow{2}{*}{Method}} & \multicolumn{2}{c}{Yelp} & \multicolumn{2}{c}{Software} & \multicolumn{2}{c}{CDs \& Vinyl} \\
 \cmidrule(lr){2-3}\cmidrule(lr){4-5}\cmidrule(lr){6-7}  
\multicolumn{1}{c}{}  & N@20 & R@20 & N@20 & R@20 & N@20 & R@20 \\
\midrule
Recformer & 4.58 & 10.48 & 10.31 & 24.27 & 2.45 & 6.19 \\
LRURec & 4.71 & 10.76 & 10.46 & 24.41 & 2.51 & 6.30 \\
COT-Rec & 5.29 & 13.06 & 11.29 & 25.47 & 3.01 & 7.36 \\
DESRec-SAS & 4.99 & 11.65 & 11.24 & 25.07 & 2.76 & 6.70 \\
DESRec-Uni & \textbf{5.78} & \textbf{13.67} & \textbf{11.64} & \textbf{26.42} & \textbf{3.26} & \textbf{7.89} \\
\bottomrule
\end{tabular}
\label{other_result}
\end{table}

\begin{figure*}[t]
\centering
\includegraphics[width=0.93\textwidth]{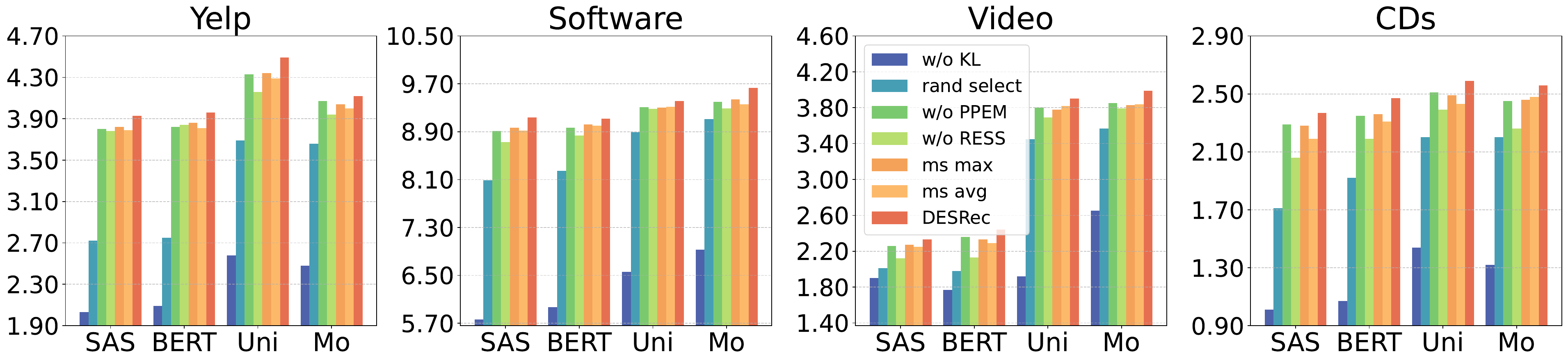}
\caption{Performance (NDCG@10) of ablation models with four backbones and four datasets. }
\label{ablation}
\end{figure*}

\begin{figure*}[t]
\centering
\includegraphics[width=0.98\textwidth]{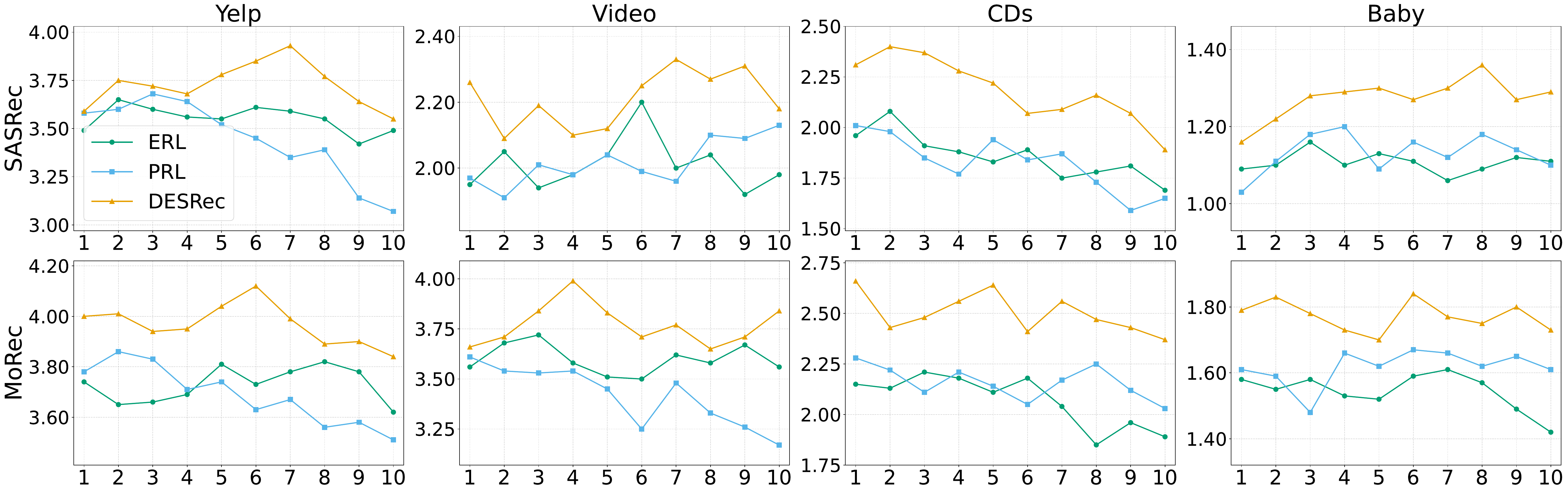}
\caption{NDCG@10 results of ERL, PRL, and DESRec with four backbones and three datasets on different reasoning steps. }
\label{parameter_step}
\end{figure*}

\section{Experimental Results}
\subsection{Main Results}
The recommendation results of DESRec and baselines on four backbones and six datasets are shown in Table~\ref{IDmodel_result}. The results of DESRec and some state-of-the-art conventional recommenders and LLM-based reasoning methods are shown in Table~\ref{other_result}. Our key findings are listed as follows:

(1) Our model outperforms all baselines on all datasets. It achieves significant improvements over the strongest baseline ReaRec in terms of SASRec backbone and NDCG@10 metric on all six datasets by 6.79\%, 5.91\%, 5.79\%, 13.94\%, 13.33\%, and 5.35\%, respectively. It demonstrates the effectiveness of DESRec in enhancing user preference modeling via our personalized training strategy and deliberate latent reasoning.
(2) Our model achieves performance improvements on both ID-based backbones and text-based backbones. It shows that DESRec can effectively enhance different backbone architectures. It also confirms that DESRec can be universally applied to various transformer-based sequential recommenders to enhance recommendation performance, demonstrating its generalizability.
(3) DESRec achieves better performance than other latent reasoning methods, and consistently outperforms them across all pre-defined reasoning steps (see Section~\ref{sec:reasoning_steps} for a detailed analysis)
It is because our method can capture the user's evolving interests through a sliding-window preference extraction and a progressive reasoning strategy. It also indicates that our model can effectively identify the optimal reasoning step and state during training and preserve the most informative state for user modeling and recommendation.
(4) Latent reasoning methods significantly outperform all backbone models. It is because the reasoning steps enable a deeper analysis of user preferences and behavioral patterns. This demonstrates the potential of latent reasoning as a promising direction for enhancing sequential recommendation architectures.
(5) Text-based models generally perform better than ID-based models. It is because they leverage not only interaction signals but also rich semantic information from item texts.
The integration of semantic information contributes to a more comprehensive understanding of user preferences.
(6) Our model achieves better performance than several state-of-the-art conventional sequential recommendation methods. This demonstrates that our approach is able to effectively enhance the performance of sequential recommendation models by leveraging multi-step implicit reasoning and capturing the optimal reasoning state.
(7) Our model achieves superior performance compared to LLM–based explicit reasoning methods by enhancing the text-based sequential recommender UniSRec. These results highlight the effectiveness of implicit reasoning for sequential recommendation and demonstrate the ability of DESRec to significantly improve the performance of traditional sequential recommendation models.

\subsection{Ablation Study}
In this section, we construct a series of ablation experiments to demonstrate the effectiveness of our model: (1) \textbf{w/o. PPEM}. We remove the progressive preference extraction component and the milestones in Section~\ref{sec:PPEM}
(2) \textbf{w/o. RESS}. We remove the optimal reasoning step selection and self-distillation component in Section~\ref{sec:RESS} and only optimize the recommendation loss on the last step.
(3) \textbf{w/o. KL}. We remove the KL loss in Equation~(\ref{KL_loss}) and only use the recommendation loss on the best step during training to optimize the model.
(4) \textbf{rand select} We replace the optimal step selection in Section~\ref{sec:step_select} with a random step selection for the subsequent computation of the recommendation loss and distillation.
(4) \textbf{milestone max}. We use max pooling to obtain the milestones in each window instead of using the last representation in Equation~(\ref{last_pool}).
(4) \textbf{milestone avg}. We use average pooling to obtain the milestones in each window instead of using the last representation in Equation~(\ref{last_pool}).

The results of the ablation study with four backbones are shown in Figure~\ref{ablation}. We can find: (1) The model \textbf{w/o. PPEM} has degraded performance, which reveals that the PPEM component can effectively capture the users' evolving interests across different periods. It helps the model to better explore users' preferences in different time phases and reason in a progressive manner for user modeling, leading to personalized recommendations. 
(2) DESRec outperforms \textbf{w/o. RESS}, which highlights the importance of RESS in identifying the optimal reasoning step for each user sequence during training and preserving the most informative states. By selecting the best reasoning step and distilling its information into the final step, RESS enhances both the stability and effectiveness of the reasoning process.
(3) The low result of \textbf{w/o KL} indicates that our model can effectively compress and preserve the optimal reasoning information into the final reasoning step, enabling more effective recommendations and demonstrating the effectiveness of the distillation process.
(4) DESRec achieves better performance than \textbf{rand select}. It reveals that DESRec can efficiently identify the optimal reasoning step during the training phase and retain its information for improved recommendation performance at test time. It further verifies the necessity and effectiveness of selecting the optimal reasoning step.
(5) DESRec consistently outperforms \textbf{milestone max} and \textbf{milestone avg}. It indicates that the last representation is more effective for constructing preference milestones, as it can better capture the user preference information within that period.

\setlength{\tabcolsep}{3.1pt}
\begin{table}[t]
\small
\centering
\caption{Results of Relative loss rates on NDCG@10, Recall@10, and cosine similarity with label items. The loss rate is computed between the results of the best and last step.}
\begin{tabular}{lcccccc}
\toprule
\multicolumn{1}{c}{\multirow{2}{*}{Method}} & \multicolumn{3}{c}{CDs \& Vinyl} & \multicolumn{3}{c}{MovieLens-1M} \\
 \cmidrule(lr){2-4}\cmidrule(lr){5-7}
\multicolumn{1}{c}{} & N@10-R & R@10-R & Sim-R & N@10-R & R@10-R & Sim-R \\
\midrule
w/o RESS-SAS & 31.79\% & 28.45\% & 33.82\% & 30.82\% & 27.16\% & 33.26\% \\
DESRec-SAS & 21.52\% & 15.30\% & 27.01\% & 24.70\% & 22.17\% & 28.17\% \\
w/o RESS-Mo & 30.67\% & 29.87\% & 20.69\% & 22.30\% & 18.53\% & 20.70\% \\
DESRec-Mo & 21.47\% & 21.23\% & 14.56\% & 14.83\% & 12.12\% & 8.28\%\\
 \bottomrule
\end{tabular}
\label{info_loss}
\end{table}

\setlength{\tabcolsep}{10pt}
\begin{table*}[t]
\small
\centering
\caption{Inference time (ms) under different reasoning step settings for SASRec, ERL, PRL, and DESRec in Yelp datasets. The results in the table are the average inference time for one test sample. Experiments are conducted on a single A800-80G GPU.}
\begin{tabular}{lccccccccccc}
\toprule
Model & step-0 & step-1 & step-2 & step-3 & step-4 & step-5 & step-6 & step-7 & step-8 & step-9 & step-10 \\
\midrule
SASRec & 0.1303 & — & — & — & — & — & — & — & — & — & — \\
ERL & — & 0.1402 & 0.1440 & 0.1536 & 0.1622 & 0.1710 & 0.1773 & 0.1886 & 0.1973 & 0.2031 & 0.2147 \\
PRL & — & 0.1367 & 0.1492 & 0.1547 & 0.1601 & 0.1708 & 0.1793 & 0.1877 & 0.1969 & 0.2069 & 0.2143 \\
DESRec & — & 0.1415 & 0.1498 & 0.1599 & 0.1667 & 0.1767 & 0.1874 & 0.1953 & 0.2063 & 0.2133 & 0.2246 \\
 \bottomrule
\end{tabular}
\label{inference_time}
\end{table*}

\begin{figure*}[t]
\centering
\includegraphics[width=\linewidth]{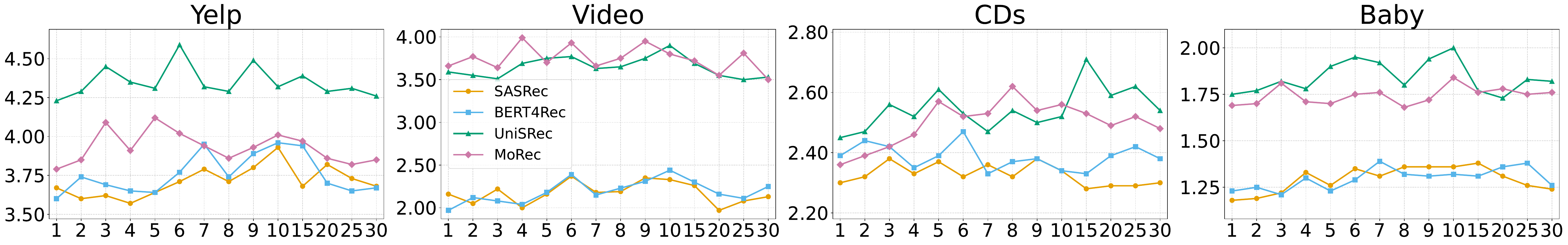}
\caption{NDCG@10 results of DESRec with four backbone models on four datasets and different window sizes.}
\label{parameter_window}
\end{figure*}

To further evaluate the fidelity of our self-distillation, we compare the results of the final step with the best step, which has the highest cosine similarity to the target label, on DESRec and \textbf{w/o RESS} using the test set. We define the relative loss rate as $(\mathrm{Best}-\mathrm{Last})/\mathrm{Best}$ and report the loss rates of NDCG@10, Recall@10, and cosine similarity between the step representations and label-item embeddings in Table~\ref{info_loss}. 
The results show that DESRec consistently achieves lower loss rates than \textbf{w/o RESS} and remains closer to the oracle best step across different backbones and datasets. This demonstrates that our self-distillation effectively captures the optimal reasoning states under different conditions and preserve the critical information into the final step, thereby improving the performance of latent reasoning and recommendation.

\subsection{Parameter Sensitivity Exploration}
The important hyperparameters in our model are the number of reasoning steps $T$ and window size $w$ (stride $D=w$) for sequence partition, and we'll analyze their impact on recommendation. 

\subsubsection{Impact of Reasoning Steps on Recommendation}~\label{sec:reasoning_steps}
We evaluate the results of NDCG@10 on six datasets under the reasoning steps $T$ ranging from 1 to 10. We compare the results of ReaRec-ERL, ReaRec-PRL, and our DESRec model. The results are shown in Figure~\ref{parameter_step}. From the results, we can observe the following:
(1) DESRec consistently outperforms ERL and PRL across the entire range of reasoning steps on all datasets and backbones. This shows that DESRec remains effective under any reasoning steps at inference and indicates the effectiveness of the PPEM and RESS components. The effectiveness and robustness of DESRec is because it can successfully distill the most informative and effective reasoning states during training, enabling more personalized and accurate recommendations.
(2) ERL and PRL models tend to achieve better performance at shallow reasoning depths (steps 2-4), but their performance degrades as the number of reasoning steps increases. In contrast, our model generally benefits from more reasoning steps, indicating its ability to effectively leverage the extra information provided by deeper reasoning.
(3) The peak performance for DESRec mostly appears in the mid-range (steps 6-9). It suggests that the model requires a sufficient depth to fully capture the user's historical behavior and preferences. However, setting the fixed inference step excessively high may occasionally introduce marginal performance fluctuations, reaffirming the importance of selecting a moderate depth for deployment.

\subsubsection{Impact of window size}
To analyze the sensitivity of DESRec to the window size, we evaluate the NDCG@10 result when $w$ is in $\{1,2,3,4,5,6,7,8,9,10,15,20,25,30\}$, respectively.
The results of four datasets with four backbones are shown in Figure~\ref{parameter_window}. The results show that DESRec achieves the best results when the window size is 5 or 10. 
A too-small window may fail to capture user preferences within a specific period and cause excessive overlap between adjacent windows. It will result in redundant or noisy signals that hinder effective reasoning.
Conversely, a too-large window may contain too many behaviors from different periods into a single window. This will introduce noise and blur the boundaries of evolving interests, which makes it difficult to track fine-grained preference changes over time.

\subsection{Inference Time Analysis}
To analyze the additional computational overhead introduced by the latent reasoning framework during the inference phase, we conduct an inference time analysis on the test set using SASRec as the backbone model. The experiments are conducted with an Intel(R) Xeon(R) Platinum 8358 CPU @ 2.60GHz and a single NVIDIA A800-80G GPU. We evaluate the backbone SASRec, ERL, PRL, and our proposed DESRec model under reasoning steps ranging from 0 to 10. Each setting is tested 20 times, and the average inference time per sample is reported in Table~\ref{inference_time}.
The results show that the latent reasoning framework introduces a certain increase in inference time, and the overhead generally grows as the number of reasoning steps increases. However, due to the use of KV cache during autoregressive reasoning, the increase in computational cost is limited. Furthermore, although the inference time increases, the actual response time per sample remains very short (e.g., 0.223 ms/sample), which is negligible compared to the overall user experience in real-world recommendation scenarios. 
These results confirm the method’s practicality and its potential for deployment in real-world applications.

\section{Conclusion}
In this work, we proposed DESRec, a deliberate latent reasoning framework that enhances sequential recommendation and user preference modeling by introducing personalized latent reasoning.
Unlike prior methods that adopt a one-size-fits-all reasoning process for all users, DESRec introduces an optimal reasoning step selection and self-distillation module that identifies the optimal reasoning depth for each user during training and preserves the most informative reasoning states into the final reasoning step for more accurate recommendation at inference.
Additionally, DESRec incorporates a progressive preference extraction mechanism that summarizes user preferences into milestones over different periods and progressively reasons over them. It facilitates modeling the evolving user preferences, especially in complex behavior sequences.
With these two components, DESRec achieves consistent improvements under any preset reasoning steps at inference.
Extensive experiments on six public datasets and four backbones demonstrate the effectiveness and generalization ability of DESRec.

\section*{Ethical Considerations}

This work evaluates recommendation models on publicly available datasets, including Yelp, Amazon, and MovieLens. These datasets are collected from real-world user-item interactions and may reflect existing biases in user behaviors and item exposure. We do not collect additional user information or utilize personally identifiable information.
Recommendation models trained on historical interactions may inherit biases such as popularity bias and unequal exposure. 
Addressing these fairness-related issues is beyond the scope of this work and can be considered in future research.



\bibliographystyle{ACM-Reference-Format}
\bibliography{sample-base}

\appendix

\end{document}